\documentclass{ar-1col}
\usepackage[numbers]{natbib}
\usepackage{bm}
\usepackage{color}
\usepackage{graphicx,epsfig}
\usepackage{xcolor}
\usepackage{amsfonts,amssymb,amsmath}
\usepackage{url}
\newcommand{\nc}{\newcommand}
\nc{\bfx}{{\bf x}}
\nc{\bfy}{{\bf y}}
\nc{\bfz}{{\bf z}}
\nc{\bfxh}{{\bf \hat{x}}}
\nc{\bfyh}{{\bf \hat{y}}}
\nc{\bfzh}{{\bf \hat{z}}}
\nc{\bfj}{{\bf j}}
\nc{\bfr}{{\bf r}}
\nc{\bfrh}{{\bf \hat{r}}}
\nc{\bfR}{{\bf R}}
\nc{\bfk}{{\bf k}}
\nc{\bfK}{{\bf K}}
\nc{\bfq}{{\bf q}}
\nc{\bfp}{{\bf p}}
\nc{\bfv}{{\bf v}}
\nc{\bfs}{{\bf s}}
\nc{\bfA}{{\bf A}}
\nc{\bfB}{{\bf B}}
\nc{\bfJ}{{\bf J}}
\nc{\bfL}{{\bf L}}
\nc{\bfS}{{\bf S}}
\nc{\bfT}{{\bf T}}
\nc{\bfY}{\bm{\mathcal{Y}}}
\nc{\bfsg}{{\bm \sigma}}
\nc{\bfta}{{\bm \tau}}
\nc{\bfvh}{{\bf \hat{v}}}
\nc{\bfqh}{{\bf \hat{q}}}
\newcommand{\mel}[3]{\langle #1 | #2 | #3 \rangle}
\newcommand{\inner}[2]{\langle #1 | #2 \rangle}
\newcommand{\avg}[1]{\langle #1 \rangle}
\newcommand{\ket}[1]{| #1 \rangle}

\def\beq{\begin{equation}}
\def\eeq{\end{equation}}
\def\beqy{\begin{eqnarray}}
\def\eeqy{\end{eqnarray}}
\jname{Xxxx. Xxx. Xxx. Xxx.}
\jvol{AA}
\jyear{YYYY}
\doi{10.1146/((please add article doi))}

\begin{document}

\markboth{King and Pastore}{Electroweak structure of light nuclei}

\title{Recent progress in the electroweak structure of light nuclei using quantum Monte Carlo methods}

\author{Garrett B. King$^1$ and Saori Pastore$^{1,2}$
\affil{$^1$Department of Physics, Washington University in Saint Louis, Saint Louis, MO 63130, USA\\ $^2$McDonnell Center for the Space Sciences at Washington University in St. Louis, MO 63130, USA}}

\begin{abstract}
Nuclei will play a prominent role in searches for physics beyond the Standard Model as the active material in experiments. In order to reliably interpret new physics signals, one needs an accurate model of the underlying nuclear dynamics. In this review, we discuss recent progress made with quantum Monte Carlo approaches for calculating the electroweak structure of light nuclei. We place particular emphasis on recent $\beta$-decay, muon capture, $0\nu\beta\beta$-decay, and electron scattering results. 
\end{abstract}

\maketitle

\tableofcontents

\section{INTRODUCTION}

Nuclear physics played a significant role in establishing the Standard Model (SM) as the theory of electroweak phenomena; however, the SM cannot describe all observed physics. Neutrino oscillations, the matter-antimatter imbalance in the universe, and the nature of dark matter all lack an adequate explanation in the SM. Electroweak phenomena involving nuclei again lie at the frontiers of physics, with the potential to improve our understanding of fundamental symmetries and resolve open questions in physics. Planned and on-going experiments-- such as precision $\beta$-decay measurements~\cite{Gonzalez-Alonso:2018omy,Falkowski:2020pma,King:2022zkz,Brodeur:2023eul}, neutrinoless double $\beta$-decay searches~\cite{Engel:2016xgb,Cirigliano:2022rmf,Agostini:2022zub}, and long-baseline neutrino oscillation experiments~\cite{Abe:2019,Acero:2019,Acciarri:2017,Aliaga:2014,Seo:2018,Abi:2020}-- aim to further constrain new physics. These efforts all share a need for an accurate theoretical understanding of nuclear electroweak structure and reactions. Thus, given the importance of these endeavors to the 2023 long-range plan for nuclear physics~\cite{lrp}, having a robust theory of nuclei with well-understood model dependencies over a wide range of kinematics is critical to reliably disentangle signals of new physics from nuclear effects. This regime encompasses the low energy domain of beta decay ($\omega\sim q \sim 0$, where $\omega$ and $q$ are the energy and momentum transfer in the process), the intermediate momentum range of neutrinoless double beta decay ($\omega \sim $ few MeVs,  $q \sim$ hundreds of MeV/c) and the quasieleastic regime ($\omega \sim$ hundreds of MeVs) relevant to precision short- and long-baseline neutrino experiments. 

A robust theory of nuclei arising from the underlying dynamics of the constituent nucleons should at best have three components: The first is a description of the interactions between nucleons in pairs, triples, and, in principle, larger groupings. Second, it entails the interaction between single nucleons, pairs and clusters of correlated of nucleons with external probes, such as electrons, photons, neutrinos and dark matter. And, finally, it needs a computational method to solve the many-body Schr\"{o}dinger equation. The reach of {\it ab initio} nuclear theory-- which has increasingly come to define systematically improvable approaches to describe nuclei and nuclear phenomena using the relevant degrees of freedom~\cite{Ekstrom:2022yea}-- has grown significantly in the last decade and there are several many-body methods available to perform calculations~\cite{Hergert:2020bxy}. In this review, we focus on quantum Monte Carlo (QMC)~\cite{Carlson:2014vla} methods, which comprise stochastic approaches to solve the many-body Schr\"{o}dinger equation for strongly correlated nucleons.

High quality nuclear interactions and currents are the input to QMC methods. Historically, phenomenological approaches were the standard for developing successful nucleon-nucleon ($NN$) interactions~\cite{Machleidt:2001,Wiringa:1994wb}. One-body electroweak transition operators from the non-relativistic reduction of the nucleon covariant currents and two-body operators arising to satisfy the continuity equation~\cite{Riska:1989bh} were the basis for studies of electroweak phenomena. 

While phenomenological models have achieved considerable success in elucidating nuclear systems and many-nucleon effects, chiral effective field theory ($\chi$EFT) has become increasingly popular.  This approach establishes a direct connection between many-nucleon interactions, electroweak currents and the underlying theory of QCD~\cite{Gross:2022hyw}. The appeal of $\chi$EFT lies in its perturbative nature, enabling a systematic expansion of interactions and currents, which, in turns, allows for a robust estimation of theoretical uncertainties. In this review, we discuss two sets of interaction and current models extensively used in QMC calculations: the Argonne $v_{18}$ (AV18) phenomenological model~\cite{Stoks:1992ja} supplemented by the Urbana IX~\cite{Wiringa:1983} or the Illinois-7~\cite{Pieper:2001ap} three-nucleon interactions with associated electroweak currents~\cite{Schiavilla:1989,Marcucci:2000,Marcucci:2005zc,Shen:2012xz}, and the Norfolk two- and three-body local chiral interaction (NV2+3)~\cite{Piarulli:2014bda,Piarulli:2016vel,Piarulli:2017dwd,Baroni:2018fdn} and associated electroweak currents~\cite{Pastore:2008ui,Pastore:2009is,Pastore:2011ip,Piarulli:2012bn,Baroni:2015uza,Baroni:2016xll,Baroni:2018fdn,Schiavilla:2018udt}. We focus on studies involving $A \le 12$ nuclei, where the nuclear many-body problem is solvable in its full complexity either exactly or within controlled approximations. Studying light systems allows for the systematic investigation of many-body effects, the assessment of various approximations used to solve the many-body problem, and the benchmarking of approximate methods applicable to the study of heavier systems of experimental interest. 

This review is structured as follows: We briefly review the QMC approaches and many-body Hamiltonian and electroweak currents in Section~\ref{sec:qmc}. Section~\ref{sec:low} reports on low-energy electroweak processes studied using QMC methods with an emphasis of recent $\beta$-decay results. Studies of process involving probes of moderate momentum transfer-- particularly, neutrinoless double $\beta$ ($0\nu\beta\beta$) decay and muon capture-- receive attention in Section~\ref{sec:moderate}. Recent advances in the study of electroweak responses to high-energy probes are covered in Section~\ref{sec:response}. Finally, we provide concluding remarks and an outlook in Section~\ref{sec:outlook}. 

\section{QUANTUM MONTE CARLO METHODS}
\label{sec:qmc}

Quantum Monte Carlo (QMC) methods and many-nucleon interactions and currents have been extensively discussed in the review articles of Refs.~\cite{Carlson:2014vla,Bacca:2014tla,Lynn:2019rdt,Gandolfi:2020pbj}. Here, we briefly highlight the salient points relevant to the calculations of electroweak observables presented in this work, directing interested readers to the aforementioned references for more details. QMC methods seek accurate solutions of the many-nucleon Schr\"{o}dinger equation
\begin{equation}
H \Psi(J^\pi;T,T_z)= E \Psi(J^\pi;T,T_z) \ ,
\end{equation}
where $\Psi(J^\pi;T,T_z)$ is a nuclear wave function with specific spin-parity 
$J^\pi$, isospin $T$, and charge state $T_z$.
The Hamiltonian has the form
\beqy \label{eq:hamiltonian} H = \sum_{i} K_i + {\sum_{i<j}} v_{ij} + \sum_{i<j<k}
V_{ijk} \ ,
\eeqy
where $K_i$ is the non-relativistic kinetic energy and $v_{ij}$ and $V_{ijk}$
are respectively two- and three-nucleon potentials. Nuclear wave functions are constructed in two steps. First, a trial variational Monte Carlo (VMC) wave function, $\ket{\Psi_T}$, that accounts for the effect of the nuclear interaction via the inclusion of correlation operators, is generated by minimizing the energy expectation value with respect to a number of variational parameters. A good variational trial function can be constructed with
\begin{equation}
   |\Psi_V\rangle =
      {\cal S} \prod_{i<j}^A
      \left[1 + U_{ij} + \sum_{k\neq i,j}^{A}\tilde{U}^{TNI}_{ijk} \right]
      |\Psi_J\rangle,
\label{eq:psit}
\end{equation}
where the Jastrow wave function $\Psi_J$ is fully antisymmetric and has the $(J^\pi;T,T_z)$ quantum numbers of the state of interest. The operators $U_{ij}$ and $\tilde{U}^{TNI}_{ijk}$ are the two- and three-body correlation. To preserve the overall anti-symmetry of the wave function, the correlation operators musted by applied symmetrically, which is accounted for with the symmetrization operator ${\mathcal S}$. 

The second step improves on $\ket{\Psi_T}$ by eliminating excited state contamination. This is accomplished in a Green’s function Monte Carlo (GFMC)~\cite{Carlson:1997qn} calculation which propagates the Schr\"{o}dinger equation in imaginary time, $\tau$, as
\begin{equation}
\ket{\Psi(\tau)} = e^{-(H-E_0)\tau}\ket{\Psi_V}\, .
\end{equation}
Since $\Psi_V$ is expandable in eigenstates of the true Hamiltonian, taking the limit $\tau \to \infty$ will remove all spurious excited state contamination. In practice, one propagates for several short imaginary time intervals until convergence is reached.

The matrix element of a transition operator $O$ should be ideally calculated in between two GFMC propagated states. Due to its prohibitive computational cost, one evaluates ``mixed estimates" where only one wave function is propagated, while the remaining one is variational, 
\begin{equation}
\avg{O(\tau)}_M = \frac{\mel{\Psi(\tau)}{H}{\Psi_V}}{\inner{\Psi(\tau)}{\Psi_V}}.
\label{eq:mixed}
\end{equation}
For diagonal transitions, a good approximation of the GFMC matrix element is then achieved under the assumption that the propagation results in a small correction to the variational state, {\it i.e.}, by assuming $\Psi(\tau) = \Psi_V + \delta\Psi(\tau)$.  Neglecting terms of order $[\delta\Psi(\tau)]^2$, one obtains 
\beq
\avg{O(\tau)} = \frac{\mel{\Psi(\tau)}{H}{\Psi(\tau)}}{\inner{\Psi(\tau)}{\Psi(\tau)}}\approx \avg{O(\tau)}_M + \left[ \avg{O(\tau)}_M - \avg{O}_{\rm VMC} \right] \, .
\label{eq:mixedestimate}
\eeq
This scheme has been generalized for off-diagonal transitions in Ref.~\cite{Pervin:2007sc}, where the expectation value of $O$ is extracted by averaging the values of $\avg{O(\tau)}$ after convergence is reached. 

\subsection{NUCLEAR HAMILTONIANS AND ELECTROWEAK CURRENTS}
\label{sec:theory}

The two- and three-nucleon potentials denoted as $v_{ij}$ and $V_{ijk}$ in Eq.~\ref{eq:hamiltonian} are inherently phenomenological. These potentials incorporate a set of parameters, encompassing underlying QCD effects, that are determined through fitting experimental data. Specifically, the two-nucleon potentials are fine-tuned to reproduce a significant number of nucleon-nucleon scattering data, along with the deuteron binding energy, achieving a $\chi^2/{\rm datum}\sim 1$. 
Typically, realistic potentials capture the long-range behavior (proportional to $1/m_\pi$ where $m_\pi$ is the pion mass) of the nuclear interaction by incorporating one-pion-exchange interaction mechanisms. Various dynamical schemes are employed to account for intermediate and short-range effects, including multiple-pion-exchange, contact interactions, or excitations of nucleons into virtual $\Delta$-isobars. 

The calculations presented in this work are based on the Argonne-$v_{18}$ (AV18) two-body interaction~\cite{Wiringa:1994wb} supplemented by the the Urbana-IX (UIX) or Illinois-7 (IL7) models of the three-nucleon force~\cite{Wiringa:1983,Pieper:2001ap}, and on the recently developed Norfolk $\chi$EFT two- and three-nucleon interactions~\cite{Piarulli:2014bda,Piarulli:2016vel,Piarulli:2017dwd,Baroni:2018fdn,Piarulli:2019cqu}.
Specifically, the AV18 two-nucleon interaction can be broken down in the following way,
\begin{equation}
v_{ij}(r) = \sum_{p=1}^{18} \left[v^p_{\pi}(r) + v^p_L(r) + v^p_S(r)\right]O^p_{ij}+ v_{ij}^{\rm EM}\, ,
\end{equation}
where $r = |\bfr_i-\bfr_j|$ is the internucleon spacing, $v^p_{\pi}$ is a long-range one-pion exchange contribution, $v^p_I$ and $v^p_S$ are phenomenological long- and short-range radial potentials, respectively, and $v^{\rm EM}_{ij}$ is the electromagnetic potential accounting for second-order Coulomb, Darwin-Foldy, vacuum polarization, and magnetic moment interaction terms. The embedded parameters were fit to $np$ and $pp$ scattering data at lab energies $E_{\rm lab} \le 350$ MeV to obtain a $\chi^2/{\rm datum}=1.1$~\cite{Wiringa:1994wb}. Three-nucleon forces constructed in combination with the AV18 interaction belong to the Urbana~\cite{Pudliner:1995wk} and Illinois series~\cite{Pieper:2001ap,doi:10.1063/1.2932280}. Both interactions
include the Fujita-Miyazawa term~\cite{Fujita:1957zz}---{\it i.e.}, a two-pion exchange
contribution involving the excitation of a virtual $\Delta$-isobar---and a short-ranged repulsive phenomenological term~\cite{Pudliner:1995wk}. The Illinois interaction adds to the Urbana one the contributions due to an S-wave two-pion exchange term plus so called ring diagrams, involving the exchanges of three pions combined with the excitation of one virtual $\Delta$-isobar~\cite{Pieper:2001ap,doi:10.1063/1.2932280}.

The Norfolk two- and three-nucleon interactions~\cite{Piarulli:2014bda,Piarulli:2016vel,Baroni:2016xll,Baroni:2018fdn,Piarulli:2019cqu}
are based on a $\chi$EFT that uses pions, nucleons and $\Delta$'s as fundamental degrees of freedom. The two-body component (denoted as NV2) is constructed up to N3LO in the chiral expansion and consists of a long-range part, $v_{ij}^{\rm L}$, mediated by one- and two-pion exchanges, and a short-range part, $v_{ij}^{\rm S}$, described in terms of contact 
interactions with strengths specified by unknown low-energy constants
(LECs). These LECs are fixed by fitting
nucleon-nucleon scattering data from the most recent and up-to-date database 
collected by the Granada group~\cite{Perez:2013jpa,Perez:2013oba,Perez:2014yla}.
The radial functions entering the NV2 potential have $1/r^n$ singularities at the origin, with $n$ reaching up to six.  These are regularized by a cutoff function that depends on a single long-range cutoff, $R_L$. Contact terms are smeared out by a Gaussian representation of the delta-function, characterized by the Gaussian parameter $R_S$. There are four model classes of the NV2 potential corresponding to different choices for the regulators and fitting energies. Model classes  (a) and (b) correspond to cutoffs [$R_L$,$R_S$]=[1.2 fm, 0.8 fm] and [1.0 fm, 0.7 fm], respectively. NV2 potentials fit with scattering data up to lab energies of 125 MeV and 200 MeV are denoted as model class I and II, respectively. 
The Norfolk three-body terms are derived consistently with the NV2 interaction up to N2LO and consist of a long-range part mediated by a two-pion exchange and a short-range component parametrized in terms of two contact interactions~\cite{vanKolck:1994yi,Epelbaum:2002vt} proportional to the LECs $c_D$ and $c_E$. Two different procedures have been adopted to constrain $c_D$ and $c_E$. In the starred (NV3$^\star$) models, $c_D$ and $c_E$ are constrained to reproduce the trinucleon binding energies along with the empirical value of the Gamow-Teller matrix element in tritium $\beta$ decay~\cite{Baroni:2018fdn}. In the unstarred models, the same LECs are fit instead to the trinucleon binding energies and the $nd$ doublet 
scattering length~\cite{Piarulli:2017dwd}. Collectively, the Norfolk two- and three-body models are denoted with `NV2+3'. 

The interaction of external probes with individual nucleons and pairs of correlated nucleons is decomposed into one- and  two-body charge and current operators as
\begin{eqnarray}
\rho    &=& \sum_i {\rho}_i({\bf q}) + \sum_{i<j} {\rho}_{ij}({\bf q}) + \dots\ , \\
\nonumber
{\bf j} &=&  \sum_i {\bf j}_i({\bf q}) + \sum_{i<j} {\bf j}_{ij}({\bf q}) +\dots  \ ,
\end{eqnarray}
where ${\bf q}$ is the momentum transferred to the nucleus.
The single nucleon charge and current operators are obtained from a non-relativistic reduction of the nucleon electroweak covariant currents~\cite{Schiavilla:1989,Marcucci:2000,Shen:2012xz}. For example, the leading order electromagnetic charge operator for point-like nucleons is simply the proton charge, while the leading order nucleon electromagnetic current consists of a convection term associated with the current generated by moving protons and spin-magnetization terms associated with the proton and neutron spins.
Studies based solely on single nucleon operators are not always sufficient to describe experimental data. For instance, two-body currents are necessary to describe electromagnetic data at both low- and high-energy regimes, including measured magnetic moments~\cite{Pastore:2012rp} and electron-nucleus cross sections~\cite{Carlson:1997qn,Lovato:2016,Lovato:2018nu}. Corrections that account for processes in which external
probes couple to pairs of correlated nucleons--described by two-body current operators--are required to explain the available data. 

In this study, we utilize two different implementations of two-nucleon electroweak currents, depending on the interaction Hamiltonian used to construct the nuclear wave functions. For calculations based on the AV18+UIX and AV18+IL7 Hamiltonians, we adopt meson-exchange currents as recently detailed in Refs.~\cite{Schiavilla:1989,Marcucci:2000,Marcucci:2005zc,Shen:2012xz}. Specifically, the two-body vector electromagnetic currents are divided into model-independent and model-dependent terms. The former are derived from the AV18, and their longitudinal components satisfy current conservation with it, ensuring their short-range behavior is consistent with that of the AV18 potential. On the other hand, the model-dependent vector electromagnetic currents are purely transverse and unconstrained by current conservation. The dominant term in this category is associated with the excitation of an intermediate $\Delta$ isobar~\cite{Schiavilla:1992sb}.
Two-body electromagnetic charge operators are inherently model dependent as they cannot be directly derived from the nucleon-nucleon interaction~\cite{Carlson:1997qn}. These operators encompass contributions from $\pi$-, $\omega$-, and $\rho$-meson exchanges, as well as $\rho\pi\gamma$ and $\omega\pi\gamma$ transition operators~\cite{Carlson:1997qn}, with pion-exchange mechanisms contributing an order of magnitude larger than the remaining terms. 
The two-body vector weak current and charge operators are derived from their electromagnetic counterparts by leveraging the conserved-vector-current hypothesis~\cite{Shen:2012xz}. Unlike the electromagnetic scenario, the axial current operator is not conserved and its two-body components cannot be directly linked to the nucleon-nucleon interaction. Among the two-body axial current operators, the leading term is associated with the excitation of $\Delta$ resonances in operators of one- and two-pion range~\cite{Shen:2012xz}.

For calculations based on the NV2+3 Hamiltonian, we employ two-body electroweak currents derived consistently with the NV2 potential~\cite{Pastore:2008ui,Pastore:2009is,Pastore:2011ip,Piarulli:2012bn,Baroni:2015uza,Baroni:2016xll,Baroni:2018fdn,Schiavilla:2018udt}. The electromagnetic charge and currents have been developed up to N3LO and N4LO , respectively. They consist of one- and two-pion range contributions along with contact terms encoding short-range dynamics. The contact terms involve two kinds of LECs: minimal and non-minimal. The minimal LECs are linked to the NV2 contact potential through current conservation, thus sharing the same LECs as the NV2 potential. The remaining LECs, including those entering a one-pion-range contribution at N3LO, in principle, could be obtained from lattice QCD calculations~\cite{Drischler:2019xuo}; however, at present, they are determined to reproduce the empirical values of the magnetic moments of $A=2$ and $3$ nuclei. For the axial charge and currents we adopt the formulation of Refs.~\cite{Baroni:2015uza,Baroni:2016xll,Baroni:2018fdn} retaining only tree-level contributions up to N3LO. These consist of one-pion-range contributions and a contact term proportional to the $c_D$ LEC (also appearing in the NV3 potential). For more in-depth information on the $\chi$EFT many-body potentials and currents, we recommend referring to the cited references for derivations based on time-ordered perturbation theory and Refs.~\cite{Krebs:2016rqz,Krebs:2019aka} for derivations based on the unitary transformation method.

\section{LOW-ENERGY ELECTROWEAK PROCESSES}
\label{sec:low}

{\it Ab initio} methods like QMC are well-controlled and highly accurate for light nuclei, making them valuable for testing nuclear models. The abundance of low-energy electroweak data available for light nuclei underscores the significance of this kinematic regime in scrutinizing models employed for the study of electroweak phenomena. For example, QMC calculations utilizing $\chi$EFT currents in combination with the AV18+IL7 potential revealed that two-body currents can contribute up to $40\%$ of the magnetic moments of nuclei with $A\le 9$~\cite{Pastore:2012rp}.
State-of-the-art magnetic moment calculations employing auxiliary field diffusion Monte Carlo approaches based on $\chi$EFT interactions and currents have provided a thorough analysis of the convergence of the chiral expansion, enabling the estimation of theoretical uncertainties~\cite{Martin:2023dhl}. This work demonstrated that the convergence is very sensitive to a consistent power counting between nuclear forces and currents, emphasizing the need for future studies to ensure that the approach is properly renormalized. It also highlighted the benefits of systematically studying low-energy observables within {\it ab initio} methods.

 Recently, there has been a significant effort to validate weak current models, driven by the potential to stringently probe BSM contributions to charge-changing weak currents using precision $\beta$-decay experiments coupled with comparably precise theoretical calculations. Promising probes of BSM physics in $\beta$-decays include the study of electron spectra, angular correlations, and tests of the unitarity of the Cabibbo-Kobayashi-Maskawa (CKM) matrix involving its first row, which currently displays a 3$\sigma$ tension with the SM expectation~\cite{Brodeur:2023eul}. The success of these efforts hinges on accurate theoretical calculations of isospin breaking effects and radiative corrections in nuclei~\cite{Hardy:2020qwl,Gorchtein:2023naa}. In response to these challenges, the Nuclear Theory for New Physics topical collaboration~\cite{ntnp} has been established with the goal of providing robust {\it ab initio} evaluations of quantities relevant to experimental endeavors. To ensure the most reliable results, a combination of models and many-body approaches for nuclei supplemented by nucleonic inputs from lattice QCD is necessary. These should accurately reproduce low-energy nuclear electroweak structure data while also consistently describing moderate-to-high energy phenomena. This approach is crucial for advancing our understanding of weak current models and their implications for precision experiments.

\subsection{$\beta$-decay in light nuclei}

\begin{figure}[h]
\includegraphics[width=3in]{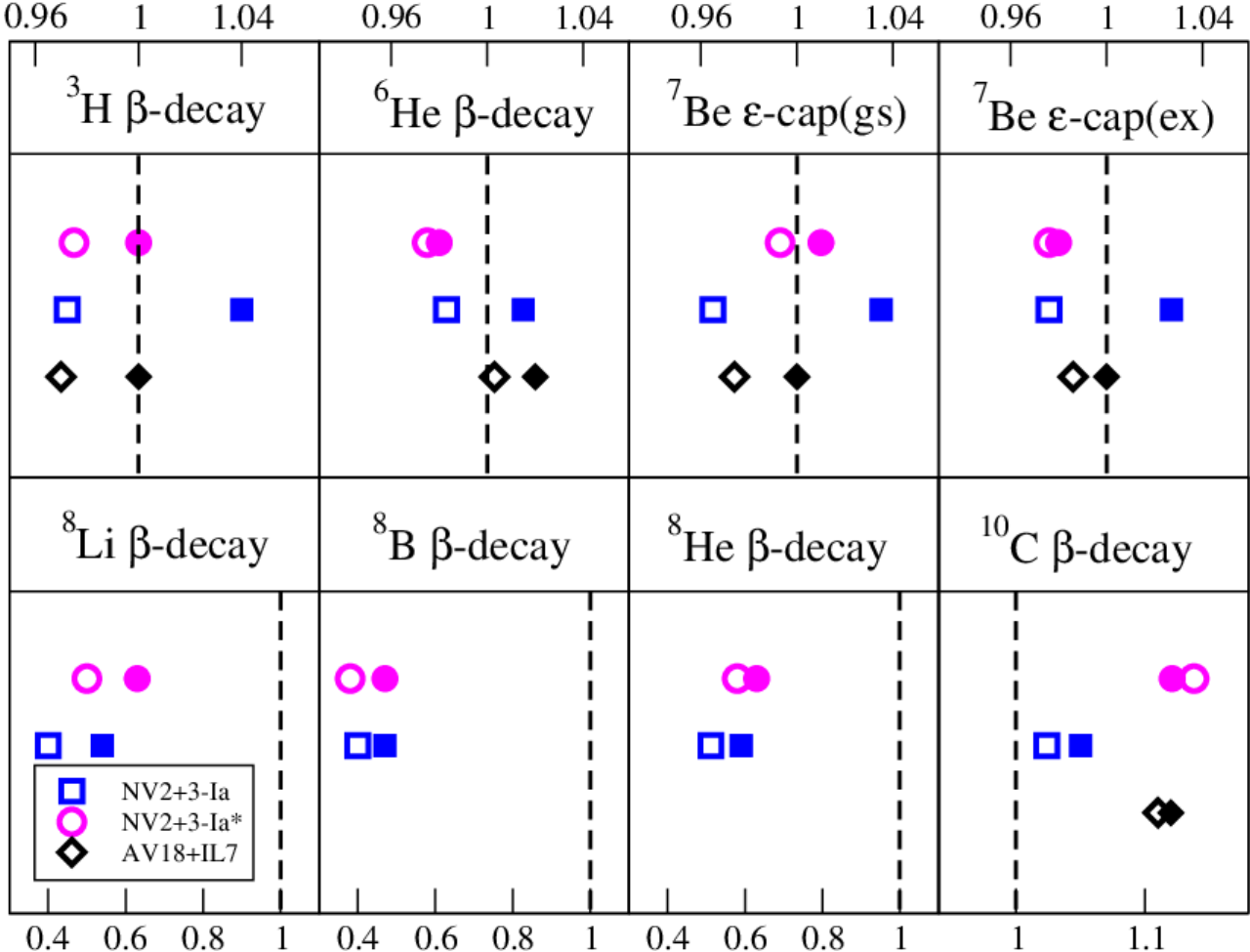}
\caption{(Color online) Ratios of $M_{\rm GT}$ calculated with GFMC to experimental values in the the $^6$He, $^7$Be, $^8$B, $^8$Be, $^8$He and $^{10}$C weak transitions. The results for $^3$H were reported in Ref.~\cite{Baroni:2018fdn}. Theory predictions corresponding to the $\chi$EFT axial current at LO (empty symbols) and up to N3LO (filled symbols) are displayed for the NV2+3-Ia (blue squares) and NV2+3-Ia$^\star$ (magenta circles) models, as well for a hybrid approach using the phenomenological AV18+IL7 Hamiltonian with $\chi$EFT currents (black diamonds). Figure reproduced from Ref.~\cite{King:2020wmp}. }
\label{fig:betaME}
\end{figure}

The investigation of $\beta$-decay has played a pivotal role in advancing our understanding of electroweak physics. Nevertheless, the study of this observable from an entirely {\it ab initio} perspective is a relatively recent undertaking. When examined from the  nuclear structure perspective, $\beta$-decay's studies have been instrumental in unveiling the role of many-body correlations and electroweak currents~\cite{Schiavilla:2002vy,Pastore:2017uwc,Gysbers:2019uyb,King:2020wmp} in the limit of momentum transfer $q$ going to zero, characteristic of this reaction. In terms of the electorweak current operators, the Gamow-Teller (GT) reduced matrix element is given as,
\beq
M_{\rm GT}= \frac{\sqrt{2J_f+1}}{g_A}\frac{\langle J_f M | j_{\pm,5}^z({\bf q}\to 0)|J_i M \rangle}{\langle J_i M, 10|J_f M \rangle}\, ,
\eeq
for an initial nucleus of angular momentum $J_i$ and projection $M$ on the spin quantization axis transitioning to a final state with the same projection but different angular momentum $J_f$. 

The long-standing issue, colloquially known as the ``$g_A$-problem", refers to the need for a systematic reduction of $M_{\rm GT}$ obtained from single-particle Shell Model calculations to match experimental data. Historically, this mismatch was resolved by an ``effective axial coupling" or ``quenching factor" that lessened the theoretical values~\cite{Chou:1993zz}. Recent {\it ab initio} computations in light and medium mass nuclei~\cite{Schiavilla:2002vy,Pastore:2017uwc,Gysbers:2019uyb, King:2020wmp} have been focused on uncovering the origin of this problem. Specifically, the work presented in Ref.~\cite{Pastore:2017uwc} reports on a set QMC calculations within the so-called ``hybrid approach" (represented in Figure~\ref{fig:betaME} with black diamonds) that combines phenomenological nuclear interactions (AV18+IL7) with $\chi$EFT transition operators. The LECs entering the $\chi$EFT currents are determined through fits based on the AV18+IL7 interactions, thereby addressing possible mismatches between the short-range dynamics embedded in the interactions and those implemented in the currents. The results of this work show that $M_{\rm GT}$ calculated using only the LO single-nucleon axial current operator and fully correlated QMC wave functions already exhibits a few percent-level agreement with the data without the need for an artificial quenching factor. The inclusion of subleading terms in the axial current enhances $M_{\rm GT}$ by 2\% to 3\%. Hence, the inclusion of correlations in the many-body wave functions emerges as the primary factor responsible for the observed quenching in light nuclei. 

\begin{figure}[h]
\includegraphics[width=5in]{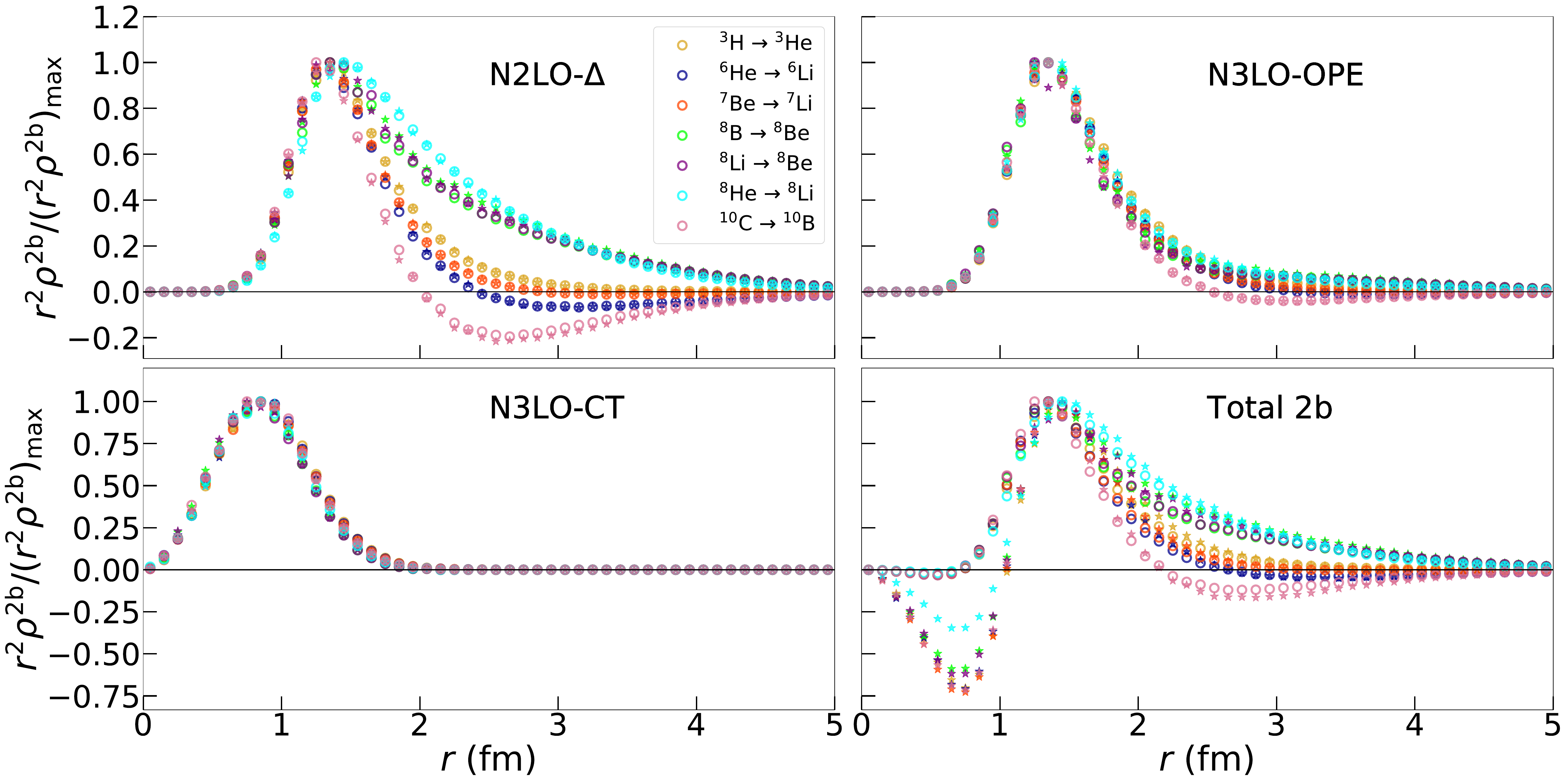}
\caption{(Color online)   Two-body transition density---see Eq.~(\ref{eq:rho2})---for 
selected transitions (represented with different colors) obtained with the NV2+3-Ia (open circles) and NV2+3-Ia$^\star$ models (filled stars). Figure reproduced from Ref.~\cite{King:2020wmp}. }
\label{fig:beta2bDensities}
\end{figure}

The quenching effect in light nuclei attributed to correlations was confirmed by a set No-Core Shell Model (NCSM) calculations utilizing $\chi$EFT interactions and currents~\cite{Gysbers:2019uyb}; however, there were discrepancies noted in the sign of the two-body current corrections for some transitions. To discern whether this discrepancy stemmed from the hybrid approach, a set of QMC calculations were carried out within a fully $\chi$EFT framework~\cite{King:2020wmp}. The authors of this study used the NV2+3 chiral interactions and associated axial currents, with particular emphasis placed on GFMC calculations based on models NV2+3-Ia and NV2+3-Ia$^\star$. As it is shown in Fig.~\ref{fig:betaME}, for all but the $A=8$ systems, the LO result is already within a few percent of the experimental datum. The subleading corrections were found to be small and additive in most cases, as was the case in the original QMC study of Ref.~\cite{Pastore:2017uwc}. However, in the $A=10$ system, the correction for the NV2+3-Ia$^\star$ model, although small, is negative. The calculations of two-body GT transition densities illuminated the origin of this sign difference. The two-body density, $\rho^{\rm 2b}$, satisfies,
\begin{equation}
M_{\rm GT}^{\rm 2b} = \int dr 4\pi r^2\rho^{\rm 2b}(r)\, ,
\label{eq:rho2}
\end{equation}
with $r$ being the internucleon distance. Fig.~\ref{fig:beta2bDensities} displays the two-body transitions densities scaled such that the point on the curve with the largest magnitude peaks at a value of one. The total $\rho^{\rm 2b}$ comprises two one-pion range contributions (N2LO-$\Delta$ and N3LO-OPE) and a strictly negative contact contribution at short-range (N3LO-CT). The N3LO-CT contribution has a universal shape across all transitions and models, with a strength that is governed by the LEC $z_0$ that is in turn proportional to the LEC $c_D$ appearing in the $3N$ force. At short distances, the densities exhibit a universal behavior arising from the fact that GT $\beta$-decay converts an $ST=10~(ST=01)$ pair to a $ST=01~(ST=10)$ pair~\cite{Schiavilla:1998je}. The associated pair densities are similar in shape to each other and exhibit a universal behavior across different nuclei within a given model~\cite{Forest:1996kp,Cruz-Torres:2019fum}. 

While the short-range behavior of $\rho^{\rm 2b}$ is the same up to an overall scaling factor, the long-range N2LO-$\Delta$ density exhibits a marked dependence on the specific transition being considered. An interplay between long- and short-range structures that cancel one another out makes the transitions sensitive to the precise overlap of the two wave functions.  For example, in the case of the $^{10}{\rm C} \to~^{10}{\rm B}$ decay, $\rho^{\rm 2b}$ clearly exhibits a node in its long-range tail. This, combined with the enhanced negative contribution from the N3LO-CT in model Ia$^\star$, results in an overall negative correction from two-body currents. Therefore, the conclusion drawn from this systematic study using different $\chi$EFT interactions is that a combination of correlations and two-body currents can lead to an overall reduction in $M_{\rm GT}$. Furthermore, two-body currents exhibit a sensitivity to different strategies implemented to determine the $3N$ LECs. This calls for further investigations to ameliorate our understanding of the $3N$ force~\cite{Piarulli2020} and its impact on several observables, including nuclear matter~\cite{Piarulli:2019pfq,Lovato:2022apd}.

In the $A=8$ systems, QMC calculations significantly underestimated the experimental data, even when accounting for the substantial correction from subleading currents. This suppression is primarily attributable to the difference in the dominant spatial symmetries of the VMC wave functions involved in the transition. This results in a reduced overlap between dominant and small components, making the transition highly sensitivity to the latter, which are poorly constrained and model dependent. Additional complications for these transitions arise from the recent findings reported in Ref.~\cite{Burkey:2022gpb} where the authors comment on a possible intruder $2^+$ state at $\sim 9$ MeV, also accessible via $\beta$-decay from $^8$Li and $^8$Be. These observations highlight the necessity for further scrutiny of these transitions within {\it ab initio} frameworks, including studies within the GFMC approach. 

The few-percent-level agreement with data in most transitions is highly encouraging, as this precision (and beyond) is essential to constrain new physics in fundamental symmetry experiments~\cite{Cirigliano:2019wao}. Precision $\beta$-decay experiments hold the potential to impose stringent constraints on BSM contributions to charge-changing weak currents~\cite{Brodeur:2023eul} and will help to better understand the 3$\sigma$ tension in the top row unitarity of the CKM matrix~\cite{Hardy:2014qxa,Gorchtein:2023naa}. This study represents an initial step toward obtaining a robust understanding with well understood model dependencies of low-energy electroweak observables within the QMC framework using the NV2+3 models. 

\subsection{$M_{GT}$ from charge exchange reactions}

GT strengths extracted from charge exchange reactions~\cite{Taddeucci:1987zz,Meharchand:2012zz,Giraud:2022ztg} offer valuable insights into nuclear models and enhance our comprehension of the microscopic structure within nuclei. Although the strong force mediates the charge-exchange reaction, it involves the same operator structures as the GT transition. The charge-exchange cross section $d\sigma/d\Omega$ can be decomposed into components with different angular momenta $\Delta L$. By fitting the cross section and extracting the $\Delta L=0$ component, the expression for $M_{\rm GT}$ is obtained through the relation:
\beq
\frac{d\sigma}{d\Omega}(q=0) \vert_{\Delta L =0} = \hat{\sigma}_{\rm GT} \frac{|M_{\rm GT}|^2}{2J_i+1}\, ,
\eeq
where $\hat{\sigma}_{\rm GT}$ is a proportionality factor depending on kinematics and the details of the nucleon-nucleus interaction used to analyze the charge exchange cross section. By comparing $M_{\rm GT}$ extracted in this way with {\it ab initio} calculations, one can learn about the wave functions of exotic nuclei with $\beta$-decays that are difficult to measure directly. 

This strategy has been adopted in Ref.~\cite{Schmitt:2022npp} where 
$M_{\rm GT}$ for decays from $^{11}$N excitations to the ground state of $^{11}$C obtained by a novel extraction using $(p,n)$ data were compared with VMC and quenched shell-model calculations.
Both theoretical approaches compared favorably with the measurement; however, the extracted $M_{\rm GT}$ values from two previous $^{11}$B charge exchange reactions-- namely, $(d,^2{\rm He})$~\cite{Ohnishi:2001kmn} and $(t,^3{\rm He})$~\cite{Daito:1998lbq}-- were discrepant. One measurement favoured an approximate isospin symmetry which would indicate relatively small continuum coupling effects, while the other was consistent with stronger effects from the continuum. VMC calculations of the mirror $\beta$-decay were found to be approximately consistent with isospin symmetry and thus aided the experimental interpretation that there is no significant difference between the low-lying states in these mirror near-dripline nuclei, as expected for $p$-shell systems. While this was a preliminary study using VMC, this work opens the door to future QMC studies of $\beta$-decay both in $A\geq 11$ nuclei and exotic systems. Further, it demonstrated the capability of the method to interface not only with $\beta$-decay, but also reactions, which will be of great importance in the FRIB era. 

\subsection{$\beta$-decay spectra and implications for BSM searches}
\label{sec:betaspectrum}

 Measurements of $\beta$-decay spectra -- {\it i.e.}, the decay rate differential in electron energy $d\Gamma/dE_e$ -- of purely GT transitions, such as $^6$He(0$^+$) $\to ^6$Li(1$^+$), aim to push constraints on physics Beyond the SM (BSM)~\cite{Cirigliano:2019wao}. On-going measurements are targeting a 0.1\% uncertainty for the $^6$He $\beta$-decay spectrum~\cite{He6CRES,Byron:2022wtr,Kanafani:2023cwv}. At this precision, deviations of the measurement from the SM prediction would indicate new physics up to the 10 TeV scale; however, this is contingent on theoretical calculations of the SM spectrum with comparable uncertainties. 

\begin{figure}[h]
\includegraphics[width=4in]{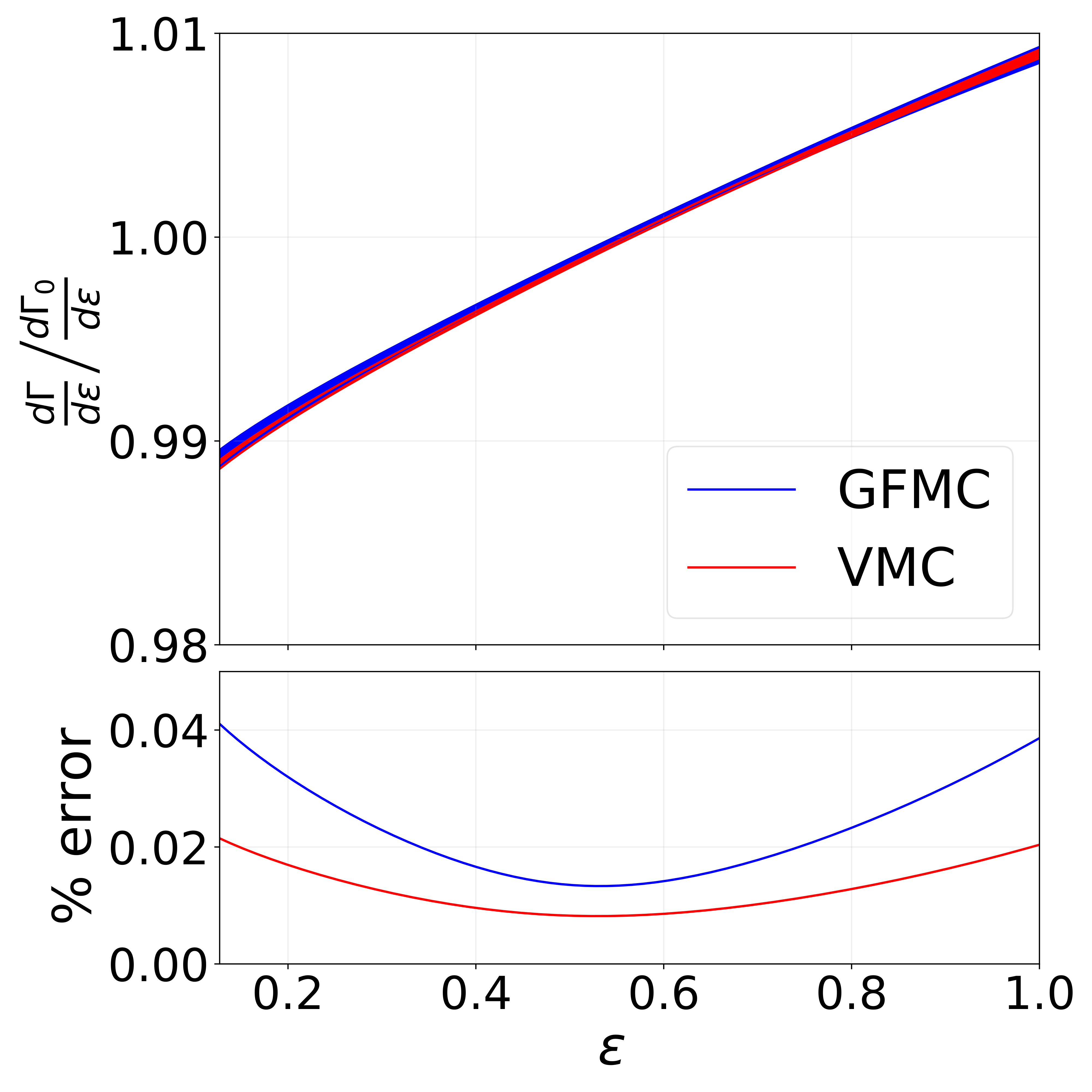}
\caption{(Color online)  The top panel displays deviations of the recoil-corrected $^6$He $\beta$-decay spectrum from the expression truncated at LO in the multipole expansion computed with VMC (red) and GFMC (blue). The NV2+3 model uncertainties of the ratios are presented in the bottom panel. Figure reproduced from Ref.~\cite{King:2021jdb}. }
\label{fig:betaSpectrum}
\end{figure}

From the theoretical standpoint, the SM differential decay rate in the typical $q \to 0$ limit is~\cite{Walecka:1995}
\beq
\frac{d\Gamma_0}{dE_e} = |M_{\rm GT}|^2\frac{G_F^2V_{ud}^2g_A^2}{2\pi^3}\omega_0(\omega_0-E_e)^2E_e^2\sqrt{1 - \frac{m_e}{E_e}}  F_0(Z,E_e)L_0(Z,E_e) S(Z,E_e)R_N(E_e)\, ,
\label{eq:q.zero.approx}
\eeq
where $G_F$ is the Fermi coupling constant, $V_{ud}$ is the CKM matrix element, $\omega_0$ is the endpoint energy of the decay, and $m_e$ is the electron mass.
In addition to $M_{GT}$, the spectrum also depends on several radiative corrections and the QMC evaluation adopted their values from Ref.~\cite{Hayen:2018}. In particular, the radiative corrections include $F_0L_0$ describing the Fermi function for a uniformly charged sphere, $S$ arising from the shielding of the nuclear charge by atomic electrons, and $R_N$ accounting for kinematic recoil corrections. Allowing for electroweak charge-changing currents beyond the SM $V-A$ structure, for this specific decay, one obtains~\cite{Jackson:1957zz} 
\beq
\frac{d\Gamma}{dE_e} = \frac{d\Gamma_0}{dE_e}\left[ 1 + b\frac{m_e}{E_e}\right]\, ,
\eeq
where $b$ is the so called ``Fierz interference term". In the SM, $b$ vanishes and thus measuring a non-zero value for this term would suggest BSM charge-changing weak currents; however, it is also possible to generate a non-zero value of $b$ accounting for recoil corrections from the small momentum transferred to the nucleus~\cite{Calaprice:1975,Holstein:1974zf} and was recently obtained in an {\it ab initio} approach using the NCSM with the NNLO$_{\rm opt}$ and NNLO$_{\rm sat}$ potentials and LO one-body transition operators~\cite{Glick-Magid:2021uwb} to obtain $b=-1.52(18)\times10^{-3}$ in the SM.

The $^6$He $\beta$ decay spectrum was recently analyzed within the QMC framework in Ref.~\cite{King:2022zkz}, where the authors accounted for the effects of two-body currents and provided a systematic analysis of the model dependence to further reduce the uncertainty on $b$. Rather than depending on $M_{\rm GT}$, the nuclear matrix element (NME) in the rate is written in terms of a standard multipole decomposition for $q\neq0$~\cite{Walecka:1995}. The multipoles can be written as NMEs calculable with QMC techniques and were evaluated at several values of $q$ in Ref.~\cite{King:2022zkz}. In order to perform a theoretical calculation with an accuracy comparable to the permille uncertainty goal of current experiments, a careful analysis of the energy and momentum scales entering the rate and ensuing kinematic corrections was carried out. In particular, the multipoles defined above were expanded in the small parameter $q/m_{\pi}$ and the expansion coefficients were extracted from interpolations of the calculated QMC values. The extracted coefficients were then inserted into the rate retaining terms up to second order in $q$ to reach an accuracy comparable to the experimental uncertainty. This was the first calculation to include second order terms in the low $q$ expansion of the multipoles that, despite their small size, help to control the size of the theoretical uncertainty on the spectrum.

Figure~\ref{fig:betaSpectrum} shows the average SM spectrum using VMC and GFMC with the associated theory uncertainties. The integrated decay rate of $808 \pm 24$ ms obtained in GFMC compares quite favorably with the recently measured rate of $807.25\pm0.16\pm0.11$ ms~\cite{Kanafani:2022tbr}. The theoretical uncertainties were obtained by using different NV2+3 potentials to investigate the impact of different choices in fitting the interaction. For example, one could estimate the $3N$ force uncertainty by evaluating NMEs in model Ia and Ia$^\star$. Then, the uncertainties on the NMEs were propagated to the expansion coefficients by fitting the multipoles with a model uncertainty at each point in $q$. Within this approach, the theoretical error of the spectrum was estimated to be below the permille level required to probe new physics in modern experiments. The SM correction $b=-1.47(3)\times10^{-3}$ obtained with QMC not only agrees with the previous {\it ab initio} evaluation, but also greatly reduces the uncertainty because of the explicit inclusion of two-body currents.

The study showed that the dominant correction to the spectrum comes from the weak magnetism term associated with the $M1$ transition mediated by the charge-changing transverse vector current. The isospin breaking corrections between the weak transition and the electromagnetic $^6$Li(0$^+$) to $^6$Li(1$^+$) transition were small enough in GFMC that one could adopt the experimental value of the latter process, further reducing the theoretical uncertainty of the spectrum. With the QMC prediction, it will be possible to distinguish signatures of charge-changing currents with tensor or pseudoscalar Lorentz structures, as well as signatures of $\sim$1 MeV sterile neutrinos. Thus, theoretical predictions for $\beta$-decay in the QMC plus $\chi$EFT framework will help to discover or place tighter constraints on BSM physics. 

\section{MODERATE-MOMENTUM ELECTROWEAK PROCESSES}
\label{sec:moderate}

Neutrinoless double beta ($0\nu\beta\beta$) decay is a process where two neutrons in a nucleus are converted into two protons with the emission of two electrons and no neutrinos, thereby violating lepton number conservation by two units. The next generation of $0\nu\beta\beta$ decay experiments holds the potential to uncover the origin of the neutrino masses and the observed matter-antimatter asymmetry in the universe~\cite{Cirigliano:2022rmf,Cirigliano:2022oqy,Engel:2016xgb,Agostini:2022zub}. The observation of this decay would imply that neutrinos are Majorana particles and serve as a probe for several lepton-number-violating (LNV) mechanisms~\cite{Cirigliano:2022rmf,Cirigliano:2022oqy}, including the standard light-Majorana neutrino exchange scheme. Regardless of the LNV decay mechanism, the interpretation of new physics hinges on an accurate understanding of the underlying nuclear dynamics. Indeed, the $0\nu\beta\beta$ decay rate is proportional to the square of an NME-- denoted by $M_{0\nu}$-- which can only be determined through theoretical computations. Specifically, the rate reads:
\beq
\Gamma_{0\nu} = G_{0\nu}|M_{0\nu}|^2\avg{m_{\beta\beta}}^2 \, , 
\eeq
where $G_{0\nu}$ is a phase space factor and $m_{\beta\beta}$ encodes new physics~\cite{Cirigliano:2022rmf,Cirigliano:2022oqy}. Assuming a light Mayorana neutrino exchange mediates the decay, this transfers momentum $q\sim 100$ MeV to the nucleus. 

Isotopes of experimental interest to $0\nu\beta\beta$ searches ({\it e.g.}, $^{48}$Ca, $^{76}$Ge, $^{82}$Se, $^{124}$Sn, $^{128}$Te, $^{130}$Te and $^{136}$Xe) are medium and heavy mass nuclei. In examining these systems, the high computational cost necessitates the usage of approximate methods to solve the nuclear many-body problem. This often entails working in truncated model spaces that may insufficiently account for correlations and many-nucleon terms in nuclear interactions. Consequently, different theoretical models can give $0\nu\beta\beta$ NMEs that vary from one another by a factor of up to two or three~\cite{Engel:2016xgb,Agostini:2022zub}. The past few years have witnessed the establishment of a combined effort from the Lattice QCD, $\chi$EFT, and {\it ab initio} nuclear theory communities aimed at delivering NMEs with minimal reliance on specific computational methods and robust estimations of theoretical uncertainties~\cite{Cirigliano:2022rmf,Cirigliano:2022oqy,Engel:2016xgb,Agostini:2022zub,Belley:2023lec}.  Simultaneously, also prompted by the Neutrinoless Double Beta Decay Topical Collaboration, the nuclear physics community is making determined efforts to advance {\it ab initio} frameworks and extend their applicability to medium mass nuclei of experimental relevance~\cite{Hergert:2020bxy}. The initial wave of calculations for the lightest nucleus that can $0\nu\beta\beta$ decay, $^{48}$Ca, came in 2021~\cite{Hergert:2020bxy,Novario:2020dmr,Yao:2019rck,Yao:2020olm,Wirth:2021pij,Coraggio:2020hwx}.

In light nuclei, the conditions for a $0\nu\beta\beta$ transition are unrealized as single $\beta$ decay is energetically allowed. Nonetheless, investigations $0\nu\beta\beta$ decay in light nuclei provide valuable insights into the dynamics of this process. The high level of accuracy achieved through {\it ab initio} calculations in these systems makes them ideal candidates for evaluating the significance of the various elements contributing to the NMEs. These encompass nucleonic form factors, many-nucleon currents, and different LNV mechanisms inducing this process. Further, when employing the $\chi$EFT framework, it becomes feasible to obtain robust estimates of the theoretical uncertainties, systematically examine the convergence of the chiral expansion, and investigate the sensitivity to the regulators. These studies offer important benchmarks for testing many-body methods that can extend to the heavier nuclei of experimental interest.

\subsection{Neutrinoless double beta decay}

QMC studies of $M_{0\nu}$ for light nuclei were first performed in Ref.~\cite{Pastore:2017ofx} using the AV18+UIX interaction. In this work, the authors examined $0\nu\beta\beta$ NMEs in $A=6-12$ nuclei induced by the standard light Mayorana neutrino exchange, and assessed the impact of subleading corrections generated by pion-neutrino loops. In the light Mayorana neutrino exchange scenario, $M_{0\nu}$ comprises three long-range contributions~\cite{Engel:2016xgb,Agostini:2022zub}. Schematically, 
\beq
M_{0\nu} = \sum_{\alpha={\rm F,GT,T}} M_{\alpha}=\sum_{\alpha={\rm F,GT,T}} \mel{J_fM_f}{O_{\alpha}}{J_iM_i}\, ,
\eeq
where the subscript refers to Fermi (F), Gamow-Teller (GT), and Tensor (T) spin-isospin structures--respectively, $\tau_a^+\tau_b^+$, $\bm{\sigma}_a \cdot \bm{\sigma}_b\,\tau_a^+\tau_b^+$, and $(3\,\bm{\sigma}_a\cdot\bfrh_{ab}\bm{\sigma}_b\cdot\bfrh_{ab}-\bm{\sigma}_a\cdot\bm{\sigma}_b)\,\tau_a^+\tau_b^+$--that enter the transition operator $O_{\alpha}$~\cite{Engel:2016xgb,Agostini:2022zub}. 

\begin{figure}[h]
\includegraphics[width=4in]{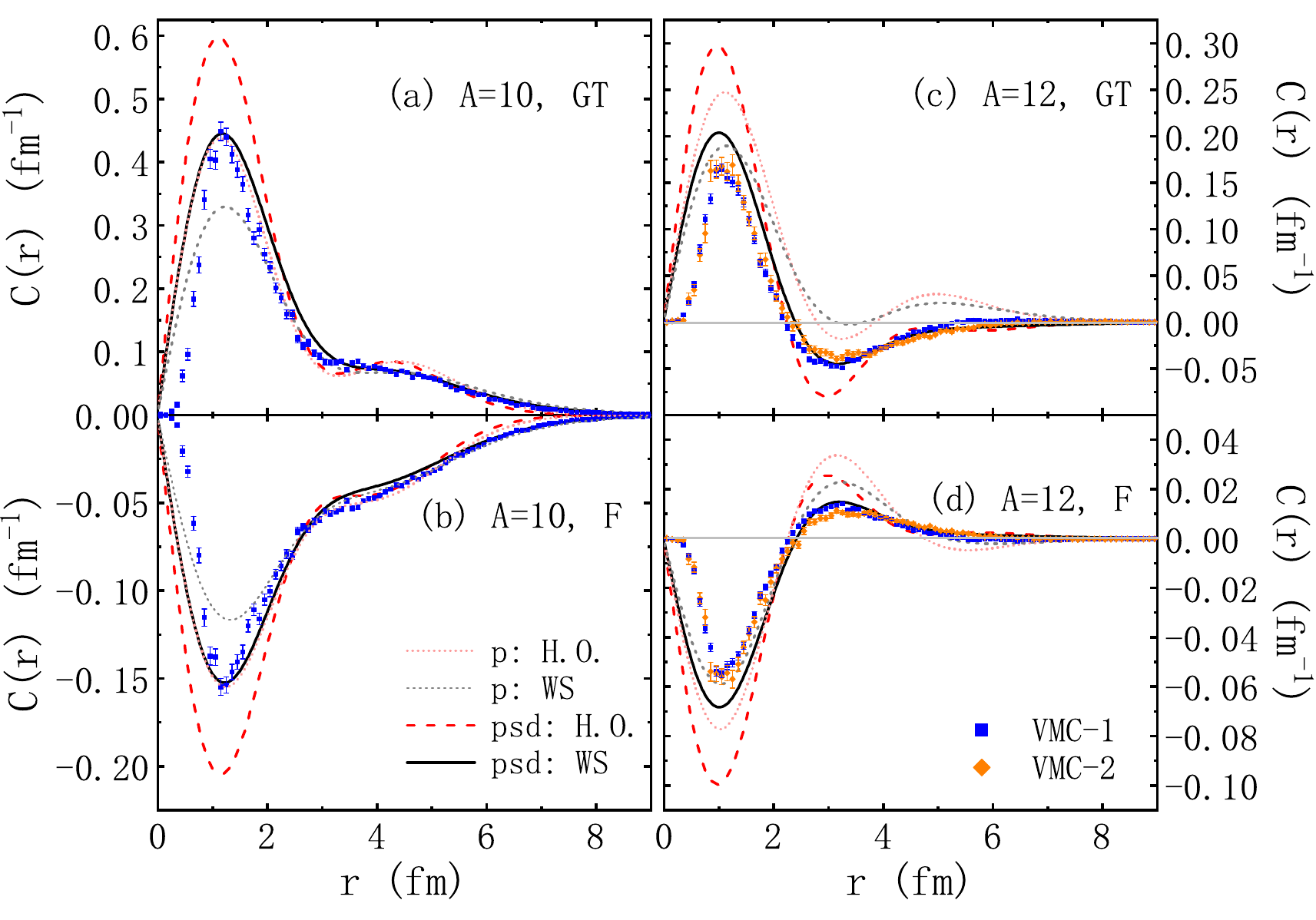}
\caption{(Color online)
Transition densities associated with F and GT operators for the $\Delta T=0$ $^{10}$Be $\rightarrow$ $^{10}$C (panels (a) and (b)) and the $\Delta T=2$ $^{12}$Be $\rightarrow$ $^{12}$C (panels (c) and (d)) transitions. The plots show the transition density $C_\alpha(r)$, where $r$ is the interparticle distance (see Eq.~\ref{eq:bbdensity}). The shell model results are for the lowest order $0\hbar\omega$ $p$-shell calculations and $psd$-shell calculations with up
to four particle excitations using two choices of radial wave functions-- Harmonic Oscillator (H.O.) and Wood-Saxson (WS)-- without short-range correlations. Figure reproduced from Ref.~\cite{Wang:2019hjy}. }
\label{fig:betabetacomparison}
\end{figure}

The authors of the study considered two types of transitions; namely, transitions where the isospin of the initial and final states is unchanged ($\Delta T=0$) and transitions where the isospin changes by two units ($\Delta T=2$). The latter transitions have direct connection to the decays of experimental interest, all of which are characterized by $\Delta T=2$. Relative to the $\Delta T=0$ NMEs, those with $\Delta T=2$ NMEs are suppressed. To understand the origin of this suppression, it is useful to analyze the transition density 
\begin{equation}
M_\alpha = \int_0^\infty  d r \, C_\alpha(r), \qquad \alpha \in \{ F, GT, T \} \, ,
\label{eq:bbdensity}
\end{equation}
where $r_{ab}$ represents the interparticale distance. 
The $\Delta T=2$ transition densities present a node which leads to non trivial cancellations, and ensuing reduction of the overall $\Delta T=2$ NME. This can be appreciated in Fig.~\ref{fig:betabetacomparison} which shows the F and GT densities for the $\Delta T=0$ $^{10}$Be $\rightarrow$ $^{10}$C (left panel) and the $\Delta T=2$ $^{12}$Be $\rightarrow$ $^{12}$C (right panel) transitions calculated using VMC wave functions (squared symbols in the figure). The study investigated the impact that correlations in the nuclear wave functions have on the calculated NMEs by artificially turning off the two-nucleon ``one-pion exchange-like"  correlation operators. Their findings indicated that, in absence of these correlations, the GT and F magnitudes increase by $\sim$10\% with respect to the correlated results. This suggests that, similarly to single-$\beta$ decay, the effect of these correlations is to quench the overall NMEs.  To date, studies in light nuclei have been performed within the QMC~\cite{Pastore:2017ofx,Cirigliano:2018hja,Wang:2019hjy,Cirigliano:2019vdj}, no-core shell model~\cite{Novario:2020dmr}, coupled-cluster~\cite{Novario:2020dmr}, in-medium similarity renormalization group~\cite{Belley:2020ejd}, in-medium generator coordinate method~\cite{Yao:2020olm,Belley:2020ejd}and importance-truncated no-core shell model~\cite{Basili:2019gvn} methods, leading to results in agreement at the 10\% level in the calculated NMEs.   

In recent years, there has been a significant theoretical effort to re-examine $0\nu\beta\beta$ from an EFT perspective~\cite{Cirigliano:2022rmf,Cirigliano:2017tvr} with the goal of establishing a systematic and controlled connection between the high-energy scale where the LNV process occurs and the low-energy scale of nuclear physics, where the phenomena would manifest. These studies indicated the need for a new leading correction to $0\nu\beta\beta$ required to obtain a finite and regulator-independent $nn \rightarrow pp$ amplitude~\cite{Cirigliano:2018hja,Cirigliano:2019vdj}. In these references, guided by renormalization arguments, the authors made the case for the introduction of a counter term that is proportional to an unknown low energy constant $g_{\nu}^{NN}$. Moreover, exploiting a relation between the $0\nu\beta\beta$ transition operators and electromagnetic operators that induce short-range corrections to the charge-independence-breaking potential~\cite{Cirigliano:2018hja,Cirigliano:2019vdj,Richardson:2021xiu}, they provided a scheme to estimate the approximate magnitude of the absolute value of $g_{\nu}^{NN}$. The latter was then used to perform QMC calculations of the associated NMEs in light nuclear systems. These studies used both phenomenological and $\chi$EFT interactions and found that the new counter term provides a sizable ($\sim 40\%$) contribution to the overall $0\nu\beta\beta$ NME~\cite{Cirigliano:2018hja,Cirigliano:2019vdj,Weiss:2021rig}. To determine the sign of $g_{\nu}^{NN}$, a new scheme-- analogous to the Cottingham formula for electromagnetic contributions to hadron masses-- was proposed in Ref.~\cite{Cirigliano:2020dmx}. This newly determined value of $g_{\nu}^{NN}$ was recently employed in calcualtions using the in-medium similarity renormalization group (IMSRG) framework~\cite{Wirth:2021pij}. In this reference, the authors demonstrated that the $g_{\nu}^{NN}$ counter term enhances the $0\nu\beta\beta$ NME by $\sim43$\% in $^{48}$Ca.

The QMC efforts directed at nuclei of experimental interest served two purposes: i) They provided benchmark calculations to guide computational methods suitable to medium mass nuclei~\cite{Wang:2019hjy} and ii) They constrained short-range dynamics in light systems and leveraged the Generalized Contact Formalism (GCF) to incorporate them into NMEs of medium mass nuclei~\cite{Weiss:2021rig}. The first effort is reported in Ref.~\cite{Wang:2019hjy} where the authors benchmarked QMC and Shell-Model calculations of $0\nu\beta\beta$ NMEs in $A=10$ and $12$ systems. QMC calculations served as a reference to determine the model dependence and uncertainties in Shell-Model approaches, thereby identifying the level of sophistication required for a robust computation of $0\nu\beta\beta$ NMEs in this framework. The study examined the impact of several factors entering Shell-Model calculations, including model space truncations, choices radial wavefunctions, and short-range correlations. All of these factors were found to be important. Specifically, a good agreement between the two approaches is achieved when larger $psd$ models space are used (as opposed to considering only $p$-shell contributions) and when Wood-Saxon (WS) single particle wave functions are adopted instead of Harmonic Oscillator (H.O.) single particle wave functions. This is apparent in Fig.~\ref{fig:betabetacomparison}, which compares Shell-Model calculations (labeled with ``$psd$ WS" and  ``$p$ H.O.")  to QMC results (represented by the squared symbols), for the $\Delta T = 0$ $^{10}$Be $\rightarrow ^{10}$C and $\Delta T = 2$ $^{12}$Be $\rightarrow ^{12}$C transitions. From the same figure, it is evident that {\it ad hoc} short-range correlations need to be included to adequately capture the short-range behavior of the densities. 

The study of Ref.~\cite{Weiss:2021rig} remedied the inadequate short-range behavior by using the GCF~\cite{Weiss:2015mba} to combine the VMC and the Shell-Model to compute $M^{0\nu}$ in nuclei with $48 \leq A \leq 136$. The GCF approach uses a separation of scales to factorize the nuclear wave function $\Psi(r_1,r_2,\ldots,r_A)$ into two parts when a pair of nucleons are close together~\cite{Weiss:2014gua}. Explicitly,
\beq
\Psi \xrightarrow[r\to0]{} \sum_{\alpha}\phi^{\alpha}(\bfr)A^{\alpha}(\bfR,\,\bfr_k \vert_{k\neq i,j})\, ,
\eeq
where $\bfr$ is the pair separation, $\bfR$ is the pair center of mass, and $\alpha$ denotes the pair quantum numbers. The zero-energy solution of the Schr\"{o}dinger Equation, $\phi^{\alpha}(\bfr)$, describes the pair dynamics at short distances and is a universal function for all nuclei that depends only on the interaction model. For $0\nu\beta\beta$ decay, one can define the ``contact" as~\cite{Weiss:2021rig}
\beq
C^{\alpha\beta}(f,i) = \frac{A(A-1)}{2}\inner{A^{\alpha}(f)}{A^{\beta}(i)}\, ,
\eeq
where $|A(i)\rangle$ and $|A(f)\rangle$ represent the residual $A-2$ initial and final states, respectively. The authors of Ref.~\cite{Weiss:2021rig} pointed out that the ratio of contacts for different nuclei are model independent; {\it i.e.},
\begin{equation}
 \frac{C^{\alpha\beta}(f_1,i_1)}{ C^{\alpha\beta}(f_2,i_2)}\bigg{|}_{\rm high~res.} = \frac{C^{\alpha\beta}(f_1,i_1)}{ C^{\alpha\beta}(f_2,i_2)}\bigg{|}_{\rm Shell-Model} \, ,
 \label{eq:gfcratio}
\end{equation}
for different $0\nu\beta\beta$ transitions. The relation in Eq.~\ref{eq:gfcratio} thus allows one to incorporate the short-range physics of the realistic, high-resolution interaction on the left into shell-model matrix elements. In this case, the high resolution interaction was the AV18+UIX. The strategy was thus i) to compute the contact for a transition accessible to both VMC with the AV18+UIX and the Shell-Model, {\it e.g.}, $C(f_2,i_2)=C(^{6}{\rm Be},^{6}{\rm He})$, and then ii) to calculate $C(f_1,i_1)\big{|}_{\rm Shell-Model}$ for a nucleus of experimental interest, and iii) to infer $C(f_1,i_1)\big{|}_{\rm VMC}$ using the relation in Eq.~\ref{eq:gfcratio}. This approach has the practical effect of adequately incorporating short-range dynamics from {\it ab initio} QMC calculations in light nuclei into Shell-Model calculations where it was previously poorly constrained.  

\begin{figure}[h]
\includegraphics[width=4in]{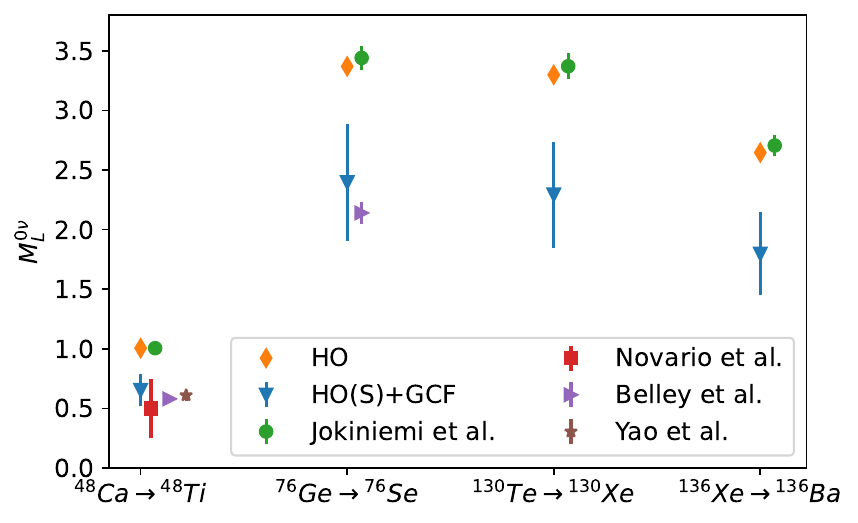}
\caption{(Color online) The light Majorana neutrino exchange contributions to $M^{0\nu}$ for $48\le A\le136$ nuclei evaluated with different many-body approaches. The GCF+SM approach incorporating short-range physics from the VMC is shown with blue triangles. Standard Shell-Model calculations are represented by orange diamonds. Other {\it ab initio} evaluations in this mass range include calculations with the shell model with a different approach to incorporate short-range correlations (green circles)~\cite{Jokiniemi:2021qqv}, coupled cluster (red squares)~\cite{Novario:2020dmr}, valence space in-medium similarity renormalization group (purple triangles)~\cite{Belley:2020ejd}, and in-medium similarity renormalization group plus generator coordinate method (brown stars)~\cite{Yao:2020olm} approaches. Figure reproduced from Ref.~\cite{Weiss:2021rig}. }
\label{fig:0nubbmes}
\end{figure}

Fig.~\ref{fig:0nubbmes} shows the light Majorana neutrino exchange contributions to $M_{0\nu}$ obtained in this formalism (labeled with `HO(S)+GCF') for various transitions. Incorporating short-range dynamics from QMC reduced the NMEs by $15\%$ to $40\%$ relative to the standard Shell-Model approach (results labeled with `HO'). This reduction brought the Shell-Model results into improved agreement with other {\it ab initio} approaches based on $\chi$EFT interactions, including the coupled cluster~\cite{Novario:2020dmr} and IMSRG~\cite{Belley:2020ejd,Yao:2020olm} methods. Both of these methods produced NMEs that are quenched relative to previous shell model analysis~\cite{Jokiniemi:2021qqv} that incorporated short-range correlations with a coupled-cluster-like approach~\cite{Simkovic:2009}. These findings confirm that a proper treatment of correlations is required and highlighted the importance of {\it ab initio} studies in light nuclei to understand many-nucleon dynamics. 

Studies of $0\nu\beta\beta$ decay NMEs have been, thus far, primarily focused on the light Majorana neutrino exchange mechanism. With the overarching goal of accurately assessing theoretical uncertainties, studies on subleading LNV contributions and their impact on $M^{0\nu}$ are now in order. These include, for example, an analysis of three-body operators~\cite{Wang:2018htk} from an EFT perspective. Light nuclei will be the ideal candidates to determine the size of these contributions and to further our understating of many-nucleon effects.

\subsection{Muon Capture}

\begin{figure}[h]
\includegraphics[width=4in]{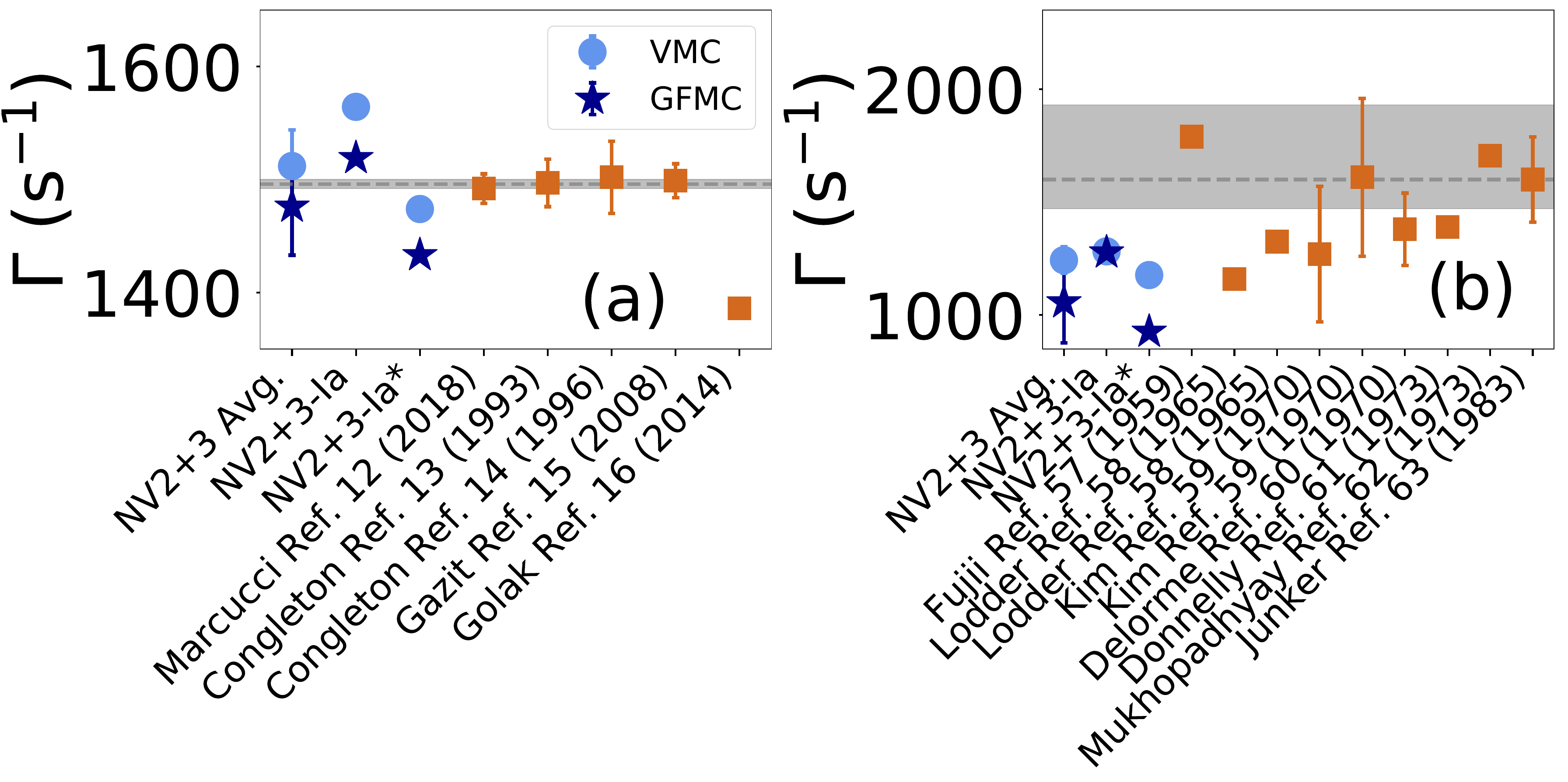}
\caption{The partial muon capture rate in (a) $^3$He and (b) $^6$Li from the the NV2+3-Ia and NV2+3-Ia$^\star$ models in VMC (light blue circle) and GFMC (dark blue star) calculations
compared with other work (orange squares)
\cite{Marcucci:2011jm,Congleton:1993epm,Congleton:1996epm,Gazit:2009mc,Golak:2014ksa,Fujii:1959,Lodder:1965,Kim:1970,Delorme:1970,Donnelly:1973,Mukhopadhyay:1973,Junker:1983}.
The experimental values (dashed gray line) and their error
(shaded region) \cite{Ackerbauer:1998,Deutsch:1968} are included for comparison with the theory predictions. Figure reproduced from Ref.~\cite{King:2022zkz}. }
\label{fig:mucap}
\end{figure}

While there are no data points with which to compare calculations of $M^{0\nu}$, muon capture-- a reaction that takes place with $q$ on the order the muon mass-- provides a way to gauge {\it ab initio} approaches by investigating a process with similar kinematics of $0\nu\beta\beta$-decay~\cite{Measday:2001,Mukhopadhyay:1977,Mukhopadhyay:1998,Kammel:2010}. In heavier systems, the process serves to study and validate many-body approaches used to study nuclei of experimental interest~\cite{Jokiniemi:2019nne,Jokiniemi:2021qqg,Simkovic:2020geo} against readily available data. In light nuclei it is a means to test electroweak currents from $\chi$EFT in calculations with controlled many-body methods beyond the more restrictive $q\to0$ limit.

 QMC methods have been recently utilized to determine partial muon capture rates in light $A=3$ and $6$ systems using the NV2+3 interaction and its consistent current operators~\cite{King:2021jdb}. In the $A=3$ system, there are muon capture measurements with sub-percent precision~\cite{Ackerbauer:1998} that are reproduced by several {\it ab initio} calculations in the $\chi$EFT approach~\cite{Marcucci:2011hh,Marcucci:2011jm,Gazit:2009mc,Golak:2014ksa}. For $A=6$, this study represented the first {\it ab initio} determination of the partial muon capture rate. 

Assuming a muon at rest in a hydrogen-like $1s$ orbital, the integrated partial capture rate from an initial state 
with angular momentum $J_i$ and projection $M_i$ to a final state $\ket{J_fM_f}$ is given by
\begin{align}
\nonumber
\Gamma &= \frac{G_F^2V_{ud}^2}{2\pi}\frac{|\psi^{\rm{av}}_{1s}|^2}{(2J_i+1)}\frac{E_{\nu}^{*2}}{{\rm recoil}}\sum_{M_f,M_i} \left\{ |\mel{J_fM_f}{\rho(E^*_{\nu}\bfzh)}{J_iM_i}|^2 + |\mel{J_fM_f}{\bfj_z(E^*_{\nu}\bfzh)}{J_iM_i}|^2  \right.  \\ \nonumber
&+ 2{\rm Re}\left[ \mel{J_fM_f}{\rho(E^*_{\nu}\bfzh)}{J_iM_i}\mel{J_fM_f}{\bfj_z(E^*_{\nu}\bfzh)}{J_iM_i}^* \right] + |\mel{J_fM_f}{\bfj_x(E^*_{\nu}\bfzh)}{J_iM_i}|^2  \\[0.25cm]
& \left. + |\mel{J_fM_f}{\bfj_y(E^*_{\nu}\bfzh)}{J_iM_i}|^2 - 2{\rm Im}\left[ \mel{J_fM_f}{\bfj_x(E^*_{\nu}\bfzh)}{J_iM_i}\mel{J_fM_f}{\bfj_y(E^*_{\nu}\bfzh)}{J_iM_i}^* \right] \right\} \, ,
\label{eq:rate}
\end{align}
where $\psi^{\rm{av}}_{1s}$ is the muon wave function averaged over the nuclear charge distribution and $E^*_{\nu}$ is the outgoing neutrino energy. The factor to account for nuclear recoil,
\begin{equation}
\frac{1}{{\rm recoil}} = \left( 1 - \frac{E^*_{\nu}}{m_i+E_0^i+m_{\mu}}\right) \, ,
\end{equation}
depends on the initial nuclear mass $m_i$, the nuclear ground state energy $E^0_i$, and the muon mass. 

In Ref.~\cite{King:2021jdb}, the authors evaluated muon capture rates with eight NV2+3 model classes and estimated the uncertainty arsing from different choices of $\chi$EFT interaction in a similar fashion to what was done in the $\beta$-decay spectrum study reviewed in Section~\ref{sec:betaspectrum}
Figure~\ref{fig:mucap} shows results for NV2+3 calculations for the $^3$He and $^6$Li ground state to ground state partial muon capture rates. For the $A=3$ calculation, rates of $1512 \pm 32~{\rm s}^{-1}$ and $1476 \pm 43~{\rm s}^{-1}$ were obtained with VMC and GFMC, respectively. The calculations were in good agreement with both the precisely measured experimental datum $1496.0 \pm 4.0 ~{\rm s}^{-1}$~\cite{Ackerbauer:1998} and previous chiral effective field theory calculations of the rate~\cite{Marcucci:2011hh,Marcucci:2011jm,Gazit:2009mc,Golak:2014ksa}. Two-body currents contributed $\sim 9\%~{\rm to}~16\%$ of the total rate in this calculation. Of the investigated sources of uncertainty, the three-body force uncertainty of $\sim 3\%$ dominated the VMC evaluation, with only modest cutoff and energy range uncertainties. 

In the case of $A=6$, the VMC and GFMC rates were $1243 \pm 59~{\rm s}^{-1}$ and $1102 \pm 176~{\rm s}^{-1}$, respectively. The two-body current contribution was smaller than in the $A=3$ transition and accounted for $3\%~{\rm to}~7\%$ of the rate. All sources of uncertainty in the $\chi$EFT approach were roughly $3\%$ each in the VMC evaluation. In the GFMC, the rather large uncertainty is due to a sharp decrease in the rate for model Ia$^\star$ after imaginary time propagation. The source of this drop was the monotonic increase in system size for $^6{\rm He}(0^+;1)$ during the GFMC propagation due to its close proximity to the $\alpha + 2n$ breakup threshold. As the current contains an $e^{i\bfq\cdot\bfr_i}$ dependence, the matrix elements is reduced because of the diffuseness of the system. It is worth noting that while the system size increases monotonically, the energy of the system and electroweak matrix element converge during the same imaginary time propagation. 

The $A=6$ rate disagreed with the experimental value of $1600^{+330}_{-129} ~{\rm s}^{-1}$~\cite{Deutsch:1968}, but fell into the range of results obtained previously with other theoretical approaches~\cite{Fujii:1959,Lodder:1965,Kim:1970,Delorme:1970,Donnelly:1973,Mukhopadhyay:1973,Junker:1983}.  
It is worth noting that the QMC calculation produces a quenched value relative to a shell-model evaluation~\cite{Donnelly:1973} and agreed with an evaluation in Ref.~\cite{Kim:1970} which treated the $^6{\rm Li}$ and $^6{\rm He}$ nuclei as elementary particles with magnetic and axial form factors extracted from experiment. Of the two calculations presented by the authors of the latter study, this result agreed with the one obtained using the Nambu formulation of the partially conserved axial current (PCAC) relation to obtain the pseudoscalar form factor, which is consistent with the induced pseudoscalar term in the $\chi$EFT weak axial current. Further studies of this observable with other methods would be valuable to validate other {\it ab initio} frameworks to reduce the uncertainty that theory contributes to experimental analyses of $M^{0\nu}$.

In addition to the calculation of muon capture to explicit final states, it is also possible to study the total muon capture rate in light nuclei. A recent QMC evaluation based on the AV18+IL7 approach was carried out in reference for the total muon capture rates of $^3$H and $^4$He~\cite{Lovato:2019mu}. In this case, rather than summing over the angular momentum projections of one final state as is done in Eq.~\ref{eq:rate}, a sum is carried out over all possible final states, including final states with nucleons in the continuum. The sum over final states is directly related to the definition of nuclear response functions,
\begin{equation}	
       R_{\alpha\beta} (q,\omega) = 
	{\overline{\sum_{M_i}}} \sum_{J_f,M_f}   \mel{J_iM_i}{j_\alpha ^\dagger ({\bf q},\omega)}{J_fM_f}	 
		 \mel{J_fM_f}{j_\beta ({\bf q},\omega)}{J_iM_i} \delta(E_f - E_i - \omega)\ ,
\label{eq:responses}
\end{equation}
where the indices $\alpha$ and $\beta$ denote the Cartesian components of the current operator. Rather than explicitly computing these highly excited states and performing the summation, it is much more practical to compute nuclear response functions via integral transform techniques. Ref.~\cite{Lovato:2019mu} evaluated the so-called ``Euclidean response" to compute the sum, an approach which will be discussed further in the following section. For $^3$He, no data were present, and a comparison was carried out with respect to the previous Fadeev calculation using only one-body currents in Ref.~\cite{Golak:2016zcw}. While the two approaches differ in the shape of the rate differential in energy, the integrated one-body capture rates of 32.4(6) s$^{-1}$ and 32.6 s$^{-1}$ for GFMC and Fadeev, respectively, were in good agreement. The inlcusion of two-body currents in the GFMC calculation changed the rate to $35.1(9)$ s$^{-1}$, accounting for $\sim 8\%$ of the total value. In the case of $^4$He, the GFMC rate was evaluated with an without the induced pseudoscalar term in the axial sector. Using the full currents, the rate of 310(12) s$^{-1}$ is agreement with the lower end of the available experimental data. Removing the induced pseudoscalar term increases the rate by $\sim14\%$ to 355(12) s$^{-1}$. Because of its possibility to more stringently constrain the pseudoscalar contribution, the authors of this study suggest that a more precise measurement of the $^4$He total capture rate would provide beneficial information that would improve the theoretical description of nuclear weak currents.  

\section{HIGH-ENERGY ELECTROWEAK PROCESSES} 
\label{sec:response}

The discovery of neutrino oscillations~\cite{sno,superk} provided compelling evidence for BSM physics. It indicated that neutrinos, assumed to be massless particles within the SM, possess non-zero masses. The neutrino experimental program is now entering a precision era~\cite{deGouvea:2022gut}, holding the potential to significantly advance our understanding of neutrino masses and their origin, CP violation, and new BSM physics. In accelerator-neutrino experiments, such as DUNE~\cite{DUNE:2022aul} and T2K~\cite{Hyper-Kamiokande:2022smq}, the detectors employ nuclei as their active material. Neutrino oscillation's parameters are extracted from the measured oscillation probabilities, which depend on the unknown incident neutrino energy. This energy is reconstructed from the final states observed in the detectors and its determination heavily relies on accurate theoretical calculations of neutrino-nucleus cross sections, along with quantifiable theoretical uncertainties. 

The QMC effort, to date, has been primarily focused on inclusive scattering processes occurring in the quasi-elastic region, characterized by energies transferred of the order of few hundreds of MeVs. Most notably, the {\it ab initio} GFMC method has been extensively used to calculate Euclidean responses induced by both electrons~\cite{Lovato:2016} and neutrinos~\cite{Lovato:2019mu} on targets up to $A=12$, the largest nuclear systems currently attainable within the GFMC. These studies show an excellent agreement between the theoretical calculations and the available data, provided that many-nucleon effects in both the nuclear interactions and currents are properly accounted for. For more details, we refer the reader to these two review articles which have recently appeared in the literature, Refs.~\cite{10.3389/fphy.2020.00116,Lovato:2023raf}.

More recently, the QMC effort has been directed towards: i) the development of  approximated {\it ab initio} methods capable of treating nuclei of experimental interest with $A>12$; ii) the extension of {\it ab initio} methods to encompass the description of exclusive scattering processes; and iii) the incorporation of relativistic kinematics for knocked out nucleons at large values of momentum transfer. To this end, two approaches have been formulated--both rooted on factorization schemes--namely, the Short-Time-Approximation (STA) and the Spectral Function (SF) approaches. The latter--recently reviewed in Refs.~\cite{10.3389/fphy.2020.00116,Lovato:2023raf}--describes the quasi elastic region by factorizing the hadronic final state in terms of a free nucleon state and $A-1$ spectator nucleons, and has been successfully extended to account for meson-production processes~\cite{Rocco:2019gfb}. 

The STA focuses on the calculation of the real-time nuclear response functions at short times, an approximation that breaks down at low  energies ($\lesssim 200$ MeV). 
Written in terms of the real time propagator, the response function is schematically given by
 \begin{equation}
  \label{eq:realtime}
     R_{\alpha}(q,\omega) = \int_{-\infty}^\infty  \frac{d t}{2 \pi} {\rm e}^{ i \left(\omega+E_i\right)   t }\, \overline{\sum_{M_i}} \mel{J_iM_i}{\mathcal{O}_\alpha^\dagger ({\bf q})\,{\rm e}^{-i H t} \mathcal{O}_\alpha ({\bf q})}{J_iM_i} \, , 
 \end{equation}
 where $|i\rangle$ is the nuclear ground state with energy $E_i$, while $\omega$ and $q$ are the energy and momentum carried by the probe, respectively. In the case of electron scattering, there are two nuclear responses, namely, the longitudinal ($\alpha=L$) induced by the charge operator, $\mathcal{O}_L=\rho$, and the transverse ($\alpha=T$) induced by the current operator, $\mathcal{O}_L={\bf j}$.
The STA fully retains many-body correlations in the ground state. However, at the vertex, where the interaction with the external probe occurs, it retains correlations and many-body currents involving at most two active particles while the remaining $A-2$ nucleons spectate. In the approximation of short times, the Hamiltonian entering the current-current correlator of Eq.~\ref{eq:realtime} includes up to two-nucleon interactions. In practice, during the scattering process, only two correlated nucleons interact with the probe via one- and two-body current operators. This allows for a correct inclusion of interference terms stemming from one- and two-body currents, which are found to be essential to explain the observed excess in the transverse electromagnetic nuclear response~\cite{Lovato:2016}.  

\begin{figure}[h]
\includegraphics[width=4in]{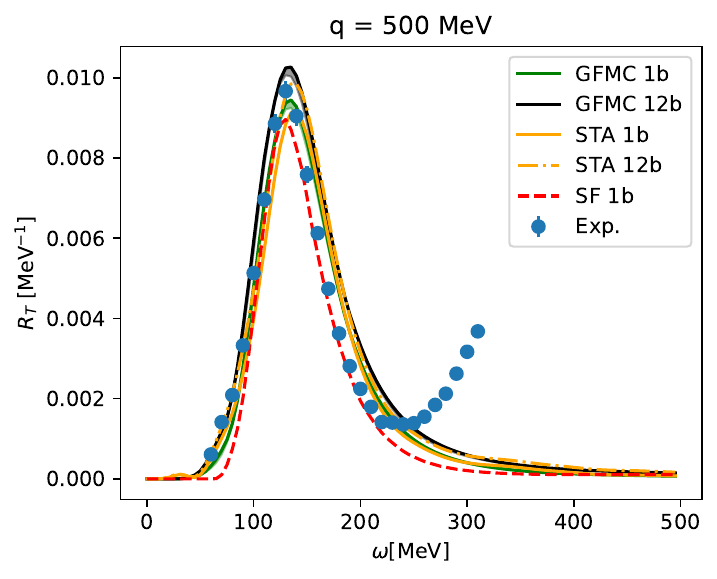}
\caption{(Color online) The transverse electron-induced response of $^3$He computed with the Euclidean response formalism using one-body only (green line) and one- and two-body currents (black line) compared with the STA one-body (solid gold) and one- and two-body (dashed gold) calculations. QMC-based spectral function formalism calculations are shown by the dashed red line and data are represented with blue circles. Figure reproduced from Ref.~\cite{Andreoli:2021cxo}. }
\label{fig:a3trans}
\end{figure}

The STA has been tested against data for electron scattering in $A=3$ and $4$ nuclei providing good agreement with both the experimental data and the GFMC calculations. As an example, Fig.~\ref{fig:a3trans} shows a comparison of Euclidean (labelled GFMC) and STA electron-induced transverse responses in $^3$He at $q=500$ MeV, obtained with one body- (labelled 1) and one- plus two-body currents (labelled 12). In this energy regime, the STA is in excellent agreement with the Euclidean response and the data; however, both the Euclidean and STA responses are missing relativistic kinematics and thus fail to reproduce the data for $q>700$ MeV. This shortcoming has been addressed in the SF approach~\cite{Benhar:2006wy,Rocco:2020jlx,Sobczyk:2023mey}, where, thanks to the factorization scheme it was possible to implement relativistic corrections, which yield to a good agreement with the data at higher kinematics~\cite{Andreoli:2021cxo}. 

\begin{figure}[h]
\includegraphics[width=4in]{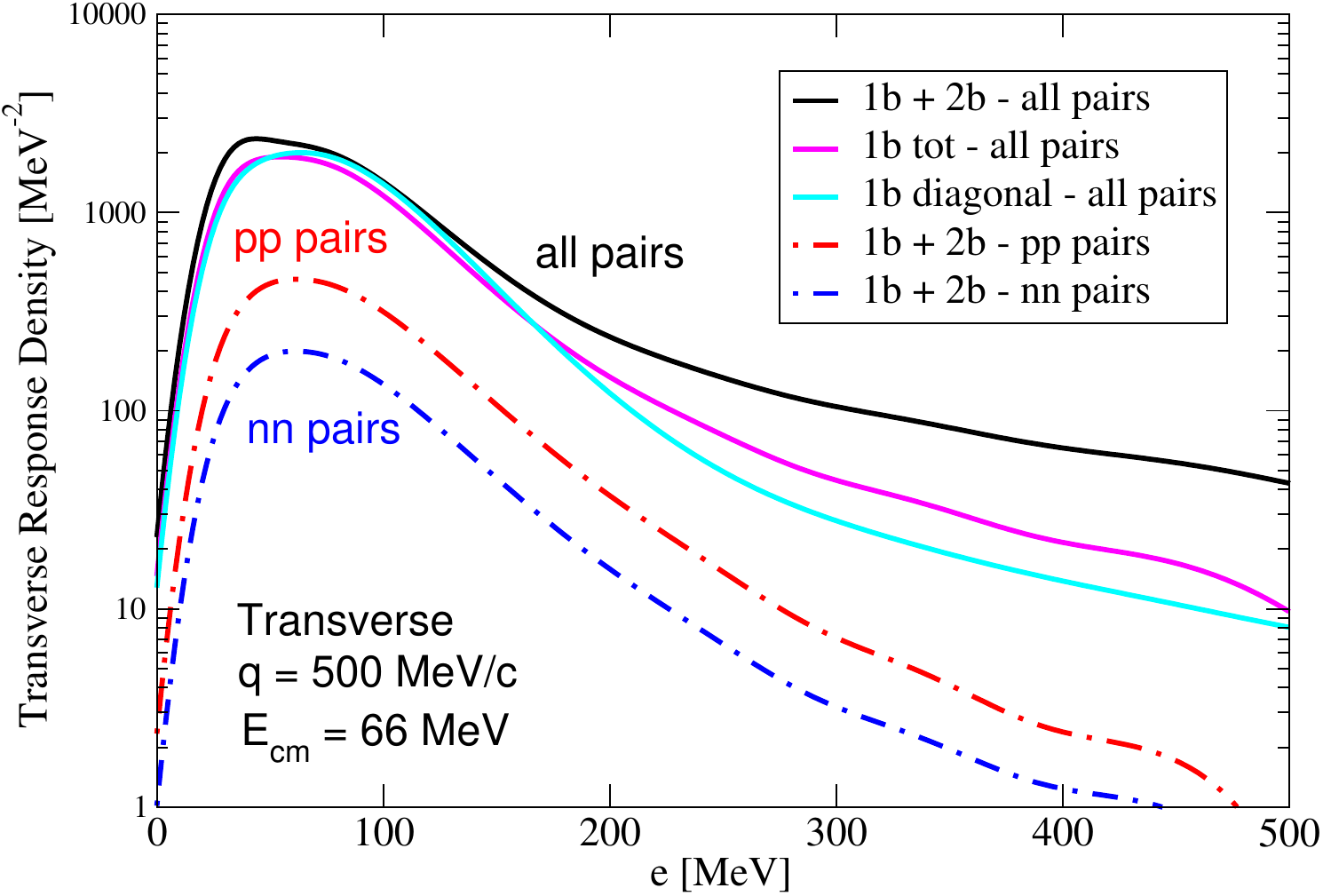}
\caption{(Color online) The slice of the three-dimensional electron-induced transverse response density of $^4$He computed with the STA sliced at $E_{CM}=q^2/(4m)$. The total one-body (pink line) and one- and two-body (black line) responses are compared with the one-body diagonal (cyan line), $pp$ one- and two-body (dashed red line), and $nn$ one- and two-body (dashed blue line) contributions. Figure reproduced from Ref.~\cite{Pastore:2019urn}. }
\label{fig:densitycut}
\end{figure}

A preliminary effort to describe exclusive processes within the STA formalism has been initiated by analyzing the response densities $D(e,E_{\rm CM})$. Here, $e$ and $E_{\rm CM}$ are the relative and center of mass energies of the struck nucleon-pair, respectively. The densities are related to the response functions via
 \beq
 R^{\rm STA}(q,\omega) = \int_0^{\infty}de \int_0^{\infty}dE_{\rm CM} \delta(\omega + E_i - e - E_{\rm CM}) D(e,E_{\rm CM})\, .
 \eeq
 Response densities provide information on the impact of two-nucleon dynamics onto scattering processes. For example, Fig.~\ref{fig:densitycut} displays the $^4$He transverse response densities at $q=500$ MeV/c in the nucleons' back-to-back kinematics ($E_{\rm CM}\sim q^2/(4m)$)~\cite{Pastore:2019urn}. In this configuration, two-nucleon currents, primarily from $np$ pairs enhances the response at large $e$. Enhancing our understanding of the  hadronic-leptonic vertex is essential for improving the energy reconstruction process, a prime component of accelerator-neutrino experiments~\cite{Barrow:2020mfy}. The STA has been validated again electron scattering data in $A\leq 4$ nuclei. Work is in progress to extend the method to $^{12}$C and incorporate relativistic effects. Furthermore, while the STA has been to date used with VMC wave function, it can be interfaced with any many-body method. Work to combine the STA with the AFDMC is being pursued with the goal of addressing nuclei with $A\sim 20$.

\section{SUMMARY AND OUTLOOK}
\label{sec:outlook}

In this review, we reported on recent progress in QMC calculations of electroweak observables in a wide range of energy and momentum transfer with emphasis on how these studies are relevant to efforts in fundamental symmetry and neutrino physics. Specifically, we presented recent calculations of beta decay and neutrinoless double decay matrix elements, beta decay spectra, muon capture rates, and lepton-nucleus response densities and functions. 

We focused on properties of light nuclei and highlighted successes and challenges that arise within the QMC framework. Light nuclei provide an optimal grounds for model validation, given the well-controlled and accurate nature of {\it ab initio} methods for these systems. These efforts unveil the dynamics and relevant physics governing effects observed in heavier systems of experimental interest. We discussed both phenomenological and $\chi$EFT based approaches and highlighted efforts towards understanding model dependencies and ensure reliable determinations of theoretical uncertainties. 

These studies contribute to the collective efforts of the nuclear physics community, employing various many-body {\it ab initio} methods combined with effective field theories, all geared towards enhancing nuclear theory and maximizing the potential of ongoing and planned experimental programs.

\section*{DISCLOSURE STATEMENT}
The authors are not aware of any affiliations, memberships, funding, or financial holdings that
might be perceived as affecting the objectivity of this review. 

\section*{ACKNOWLEDGMENTS}

We thank M. Piarulli, L. Andreoli, J. Carlson, V. Cirigliano, S. Gandolfi, A. Hayes, A. Lovato, J. Menendez, E. Mereghetti, N. Rocco, R. Schiavilla, I. Tews,  X. Wang, R. Weiss,  and R. Wiringa for their permission to reproduce the figures in this work and for valuable feedback at various stages of the manuscript.
This work is supported by the U.S.~Department of Energy under contract DE-SC0021027 (G.~B.~K. and S.~P.) and the U.S.\ DOE NNSA Stewardship Science Graduate Fellowship under Cooperative Agreement DE-NA0003960 (G.~B.~K.). We thank the Nuclear Theory for New Physics Topical Collaboration, supported by the U.S.~Department of Energy under contract DE-SC0023663, for fostering dynamic collaborations.

%

\bibliographystyle{ar-style5}
\bibliography{beta}
 
\end{document}